\documentclass[aps,prl,twocolumn]{revtex4}

\usepackage{amsmath,bm,epsfig}
\def \ed {\end{document}}
\def\Fbox#1{\vskip1ex\hbox to 8.5cm{\hfil\fboxsep0.3cm\fbox{%
  \parbox{8.0cm}{#1}}\hfil}\vskip1ex\noindent}  %%  {TEXT} in BOX

\let\*\cdot
\def\<{\left\langle} \def\>{\right\rangle} \def\({\left(} \def\){\right)}
 \let\~\widetilde \let\^\widehat 
   
  \def\1{\bm1} 
 
\newcommand{\B}[1]{{\bm{#1}}}%% Bold Roman & Greek Lower & Upper Case
\newcommand{\C}[1]{{\mathcal{#1}}}    %%   Calligrapfic Upper case
\def\BE{\begin{equation}}\def\EE{\end{equation}}
\def\BEA{\begin{eqnarray}}\def\EEA{\end{eqnarray}}
\def\BSE{\begin{subequations}}\def\ESE{\end{subequations}}

\renewcommand{\sb}[1]{_{\text {#1}}}  %% sub-   for lower case
\newcommand{\Sp}[1]{^{^{\text {#1}}}} %% Super- for Upper case
  \def\Sb#1{_{\scriptscriptstyle\rm{#1}}}
\newcommand{\eq}[1]{(\ref{#1})}%%  requires \eq{label}
\newcommand{\Eq}[1]{Eq.~(\ref{#1})}%%  requires \eq{label}
\newcommand{\Eqs}[1]{Eqs.~(\ref{#1})}%%  requires \eq{label}
\newcommand{\ve}{\varepsilon}

\def\Ret {\mbox{Re}_\tau}
\let \= \equiv  
\def\({\left(} \def\){\right)}
 \def \[ {\left [} \def \] {\right ]}
   \def\Sp#1{^{\scriptscriptstyle\rm{#1}}}
    \let\^\widehat
  \let\-\overline

\begin{document}

\title{Universal Model of Finite-Reynolds Number Turbulent Flow  in Channels and Pipes }

\author{Victor S. L'vov, Itamar Procaccia and Oleksii Rudenko}
\affiliation{Department  of Chemical Physics, The Weizmann Institute
of Science, Rehovot 76100, Israel}
\begin{abstract}
In this Letter we suggest a simple and physically transparent analytical model of pressure driven turbulent wall-bounded flows at high but finite Reynolds numbers Re. The model provides an accurate quantitative description of the profiles of the mean-velocity and Reynolds-stresses (second order correlations of velocity fluctuations) throughout the entire channel or pipe, for a wide range of Re, using only three Re-independent parameters. The model sheds light on the long-standing controversy between supporters of the century-old log-law theory of von-K\`arm\`an and Prandtl and proposers of a newer theory promoting power laws to describe the intermediate region of the mean velocity profile.
\end{abstract}
\maketitle

An important challenge in wall-bounded Newtonian turbulence is the description of the profiles of the mean velocity and second order correlation functions of turbulent-velocity fluctuations throughout the entire channel or pipe at relatively high but finite Reynolds numbers. To understand the issue, focus on a channel of width 2$L$ between its parallel walls, where the incompressible fluid velocity $\bm U(\bm r,t)$  is decomposed into its average (over time) and a fluctuating part
$$
 \bm U(\bm r,t) = \bm{V}(\bm r)  + \bm u(\bm r,t) \ , \ \bm V(\bm
r) \equiv \langle \bm U(\bm r,t) \rangle$$.
 In a stationary plane channel flow with a constant pressure gradient
$p'\equiv -\partial p/\partial x$ the only component of the mean velocity $\B V$ is the
stream-wise component $V_x\=V$ that depends on wall normal
direction $z$ only. Near the wall the mean velocity profiles for different Reynolds
numbers exhibit data collapse once presented in wall units,
 where the Reynolds number $\Ret$, the normalized
distance from the wall $z^+$ and the normalized mean velocity
$V^+(z^+)$  are defined (for channels) by
$$
\Ret \equiv {L\sqrt{\mathstrut p' L}}/{\nu}\ , \  z^+
\equiv {z \Ret }/{L} \ , \  V^+ \equiv
{V}/{\sqrt{\mathstrut p'L}}$$.

\begin{figure*}
\includegraphics[width=0.48 \textwidth]{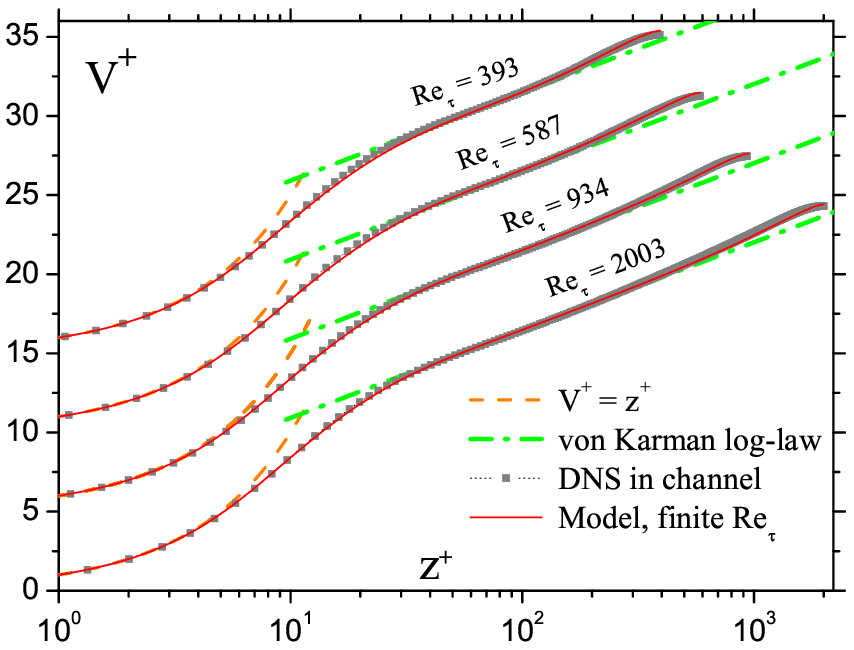}~~~
\includegraphics[width=0.48  \textwidth]{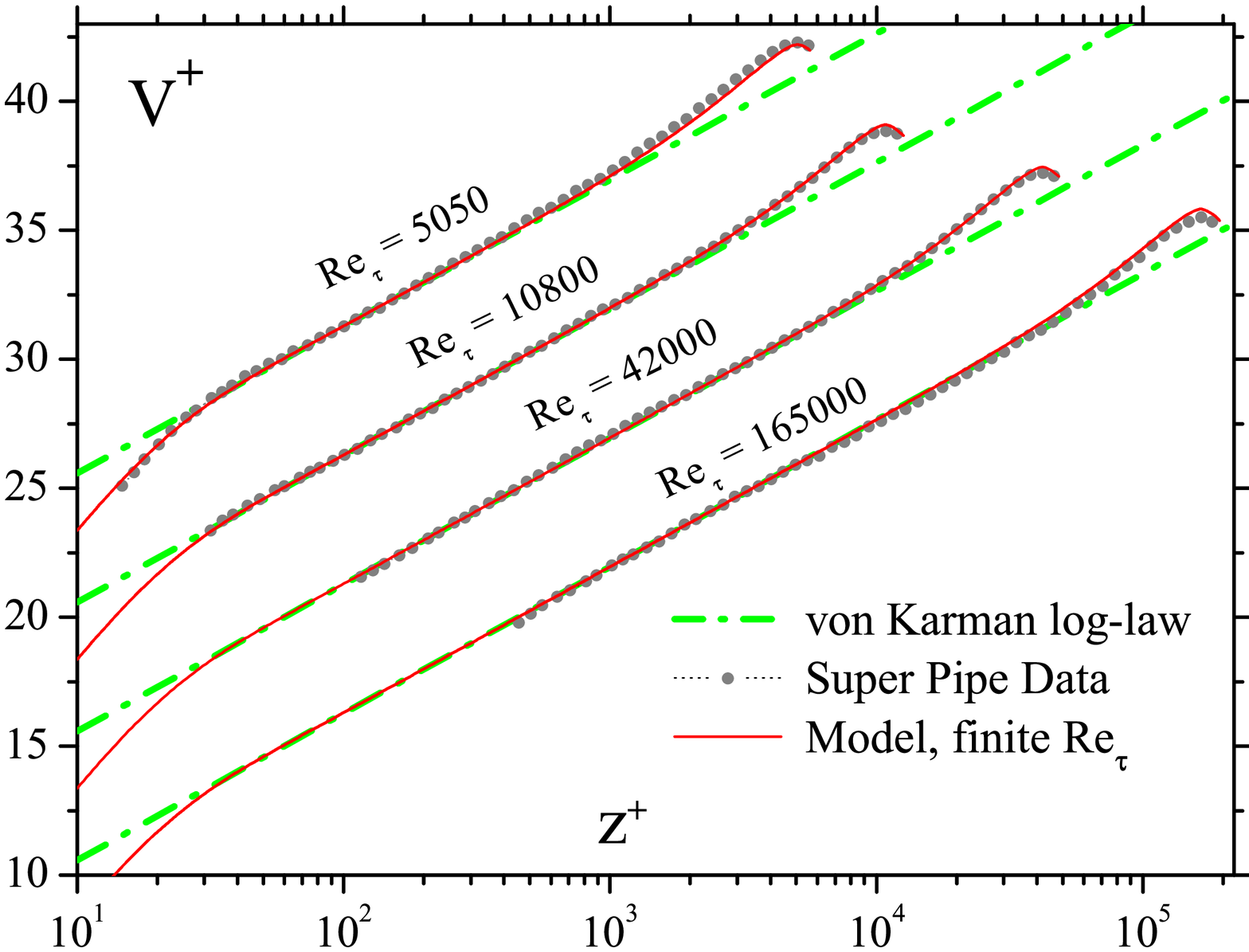}\\ ~\\
\includegraphics[width=0.51 \textwidth]{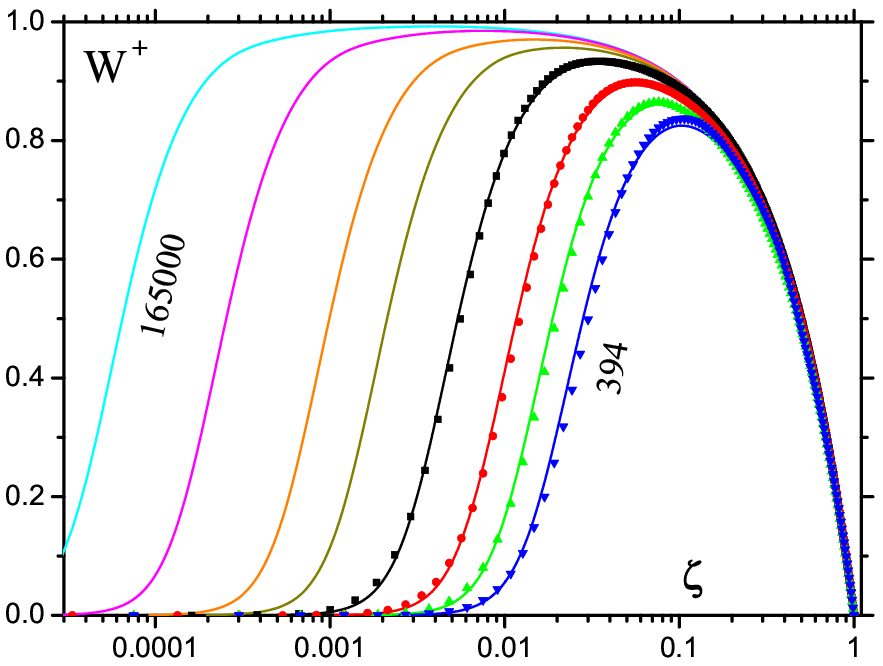}
\caption{Color online. {\bf Left and Right upper panels}: comparison of the theoretical mean velocity profiles (red solid lines) at different values of $\Ret$ with the DNS data  for the channel flow~\cite{Moser,DNS} (Left panel,  grey squares; model with $\ell\sb{buf}=49,\ \kappa=0.415, \ \ell\sb s=0.311$) and with the experimental Super-Pipe data~\cite{princeton} (middle panel, grey circles; model with $\ell\sb{buf}=46,\ \kappa=0.405, \ \ell\sb s=0.275$). In orange dashed line we plot the viscous solution $V^+=z^+$. In green dashed dotted line we present the von-K\`arm\`an log-law. Note that the theoretical predictions with three $\Ret$-independent parameters fits the data throughout the channel and pipe, from the viscous scale, through the buffer layer, the log-layer and the wake. For clarity the plots are shifted vertically by five units. {\bf Lower panel}: The Reynolds-stress profiles (solid lines) at  $\Ret$ from 394 to 2003 (in channel) and from 5050 to 165,000 (in pipe)  in comparison with available DNS data (dots) for the channel.}
\label{profiles}
\end{figure*}

The classical theory of Prandtl and von-K\`arm\`an for infinitely large $\Ret$ is based on the assumption that the single characteristic scale in the problem  is proportional to the distance from the (nearest) wall.   It leads to the celebrated  von-K\`arm\`an log-law~\cite{Pope}
\begin{equation}
 V^+(z^+) =  \kappa^{-1}\ln (z^+)  +B \,, \label{loglaw}
 \end{equation}
 which serves as a basis for the parametrization of turbulent flows near a wall in many engineering applications.
On the face of it this law agrees with the data (see, e.g. Fig.~\ref{profiles}) for relatively large $z^+$, say for $z^+>100$, giving $\kappa\sim 0.4$ and $B\sim 5$.  The range of  validly of the   log-law is definitely restricted   by the requirement $\zeta\ll 1$, where  $\zeta\=z/L$ (channel) ore $\zeta\=r/R$ (Pipe of radius $R$).  For $\zeta\sim 1$ the global geometry   becomes important leading to unavoidable
deviations of $V^+ (\zeta)$ from the log-law~\eq{loglaw}, known as {\it the wake}.

%%%%%%%%%%%%%%%%%%%%%%%%%%%%%%%%%%%%%%%%%%%%%%%%% FIG 1 %%%%%%%%%%%%%%%%%

The  problem is that for finite $\Ret$ the corrections to the  log-law~\eq{loglaw} are in   powers of $\varepsilon\equiv 1/\ln \Ret$~\cite{93Bar} and
definitely cannot be neglected for the currently largest available direct numerical simulation (DNS) of  channel flows ($\Ret=2003$~\cite{DNS} , giving $\ve\approx 0.13$). Even for $\Ret$  approaching $500,000$  as in the Princeton Superpipe experiment~\cite{princeton}, $\ve\approx 0.08$. This opens a Pandora box with various possibilities to revise the log-law~\eq{loglaw} and to replace it, as  was suggested in~\cite{93Bar},  by a power law
 \begin{equation}\label{BC}
V^+(z^+) = C(\Ret) (z^+)^{\gamma(\Ret)}\ .
\end{equation}
Here both    $C(\Ret) $ and   $\gamma(\Ret)$ were represented as asymptotic series expansions in $\varepsilon$.  The relative complexity of this proposition compared to the simplicity of Eq.~(\ref{loglaw}) resulted in a less than enthusiastic response in the fluid mechanics community \cite{98SZ}, leading to a rather fierce controversy between the log-law camp and the power-law camp. Various attempts~\cite{princeton,93Bar,98SZ,WKG,RLP,Nagib} to validate the log-law~\eq{loglaw} or the alternative power-law~\eq{BC} were based on extensive analysis of experimental data used to fit  the velocity profiles as a formal expansion in inverse powers of $\ve$ or as composite expansions in both $z^+$ and $\zeta$. Note however that in the excellent fits presented, say in
\cite{Nagib}, one uses four adjustable parameters for each function.

In this Letter we propose a complementary approach to this issue which will finally use only three  $\Ret$-independent {\em universal} parameters which will be used for all the functions discussed. First we ask what could be missed in the textbook derivations of the classical log-law~\eq{loglaw} which may lead to different velocity profile [including possibly the power   law~\eq{BC}]? Our answer is: the mean turbulent velocity profile in the entire  channel or pipe can be described within the traditional approach if one realizes how  the characteristic length-scale, which has physical meaning of the size of energy containing eddies $\ell$,  depends on the position in the flow. Simple scaling near the wall, $\ell ^+=\kappa z^+$, leads to the log-law~\eq{loglaw}.   The alternative suggestion of~\cite{93Bar}, $\ell^+ \propto (z^+)^{\alpha(\Ret)}$, leads to alternative power-law~\eq{BC}. We see no physical reason why $\ell$ should behave in either manner. Instead, we propose that the eddy size $\ell$ should be about  $z$ for $z\ll L$, and saturate at some level  $\ell\sb s  \lesssim L$ approaching the center-line, where the effect of the opposite wall is felt.  Our analysis of DNS data provides a strong support to this idea, allowing us to get, within the traditional (second-order) closure procedure, a quantitative description of the mean shear,  $S(z)=d V(z)/dz$,   the kinetic energy density (per unit mass), $K(z)\= \< |\B u|^2\>/2$, and the tangential Reynolds stress, $W(z)\= -\< u_x u_z\>$,   in the entire flow and in a wide region of $\Ret$, using only three $\Ret$-independent parameters, $\kappa$, $B$ and $\ell\sb s$ ($\ell\sb s\approx 0.311\,  L$ for the channel and $\ell\sb s\approx 0.275 \,L$  for the pipe).

\noindent
\textbf{The closure model} should relate three objects: $S^+$, $K^+$ and $W^+$.
The first (exact) relation between these objects follows from the Navier-Stokes equation for the mean velocity, being the mechanical balance between the
  momentum generated at distance $z$ from the wall, i.e. $p'(L-z)$,
  and the momentum transferred to the wall by kinematic viscosity and turbulent transport. In physical and wall units it has the form:
   \begin{equation}
 \nu S + W= p'(L-z) \!\Rightarrow\! S^+   +W^+  =1-\zeta \ , \quad \zeta\= z/L\ . \label{mom}
  \end{equation}
  Already in 1877 Boussinesq attempted to close this equation by introducing the notion
  of turbulent viscosity $\nu_{_{\rm T}}$, writing $W = \nu_{_{\rm T}} S$ \cite{Bus}. Estimating $\nu_{_{\rm T}}$ as  $\kappa_{_W}\ell_{_W}\sqrt K $, one finishes with the closure  $W^+ =  \kappa_{_W} \ell^+_{_W}\sqrt{K^+} S^+$. Here $\ell_{_W}$ is a $\zeta$-dependent characteristic scale of energy containing eddies, determining the nonlinear dissipation of $W$, and $\kappa_{_W}$ is a constant introduced here for convenience. A more careful analysis of the balance equation for $W$ (see Ref.~\cite{06LPR} and Appendix) that includes the viscous dissipation of $W$,  leads to a  somewhat more invloved closure for $W$ in a form involving an additional {\em universal}, $\Ret$-independent dimensionless function of $z^+$:
  \begin{equation}\label{W}
   r_{_W} W^+  \approx   \kappa_{_W}  \ell^+_{_W}\sqrt{K^+} S^+,\
  r_{_W}(z^+)\= \Big (1+ \frac{ {\ell\sb
{buf}^+ }^6 }{ {z^+}^6 } \Big )^{1/6}    .
 \end{equation}
Here  $\ell\sb {buf}^+ \approx 49$ is a $\Ret$-independent length that plays a role of the crossover scale (in wall units) between the  buffer and log-law region. In this form,  $\ell_{_W}(\zeta)\propto z$ near the wall, and the choice $\kappa_{_W}\approx 0.20$ ensures that $\lim _{\zeta\to 0} \ell_{_W}(\zeta) =\zeta$.

 %%%%%%%%%%%%%%%%
\begin{figure*}
\includegraphics[width=0.48 \textwidth]{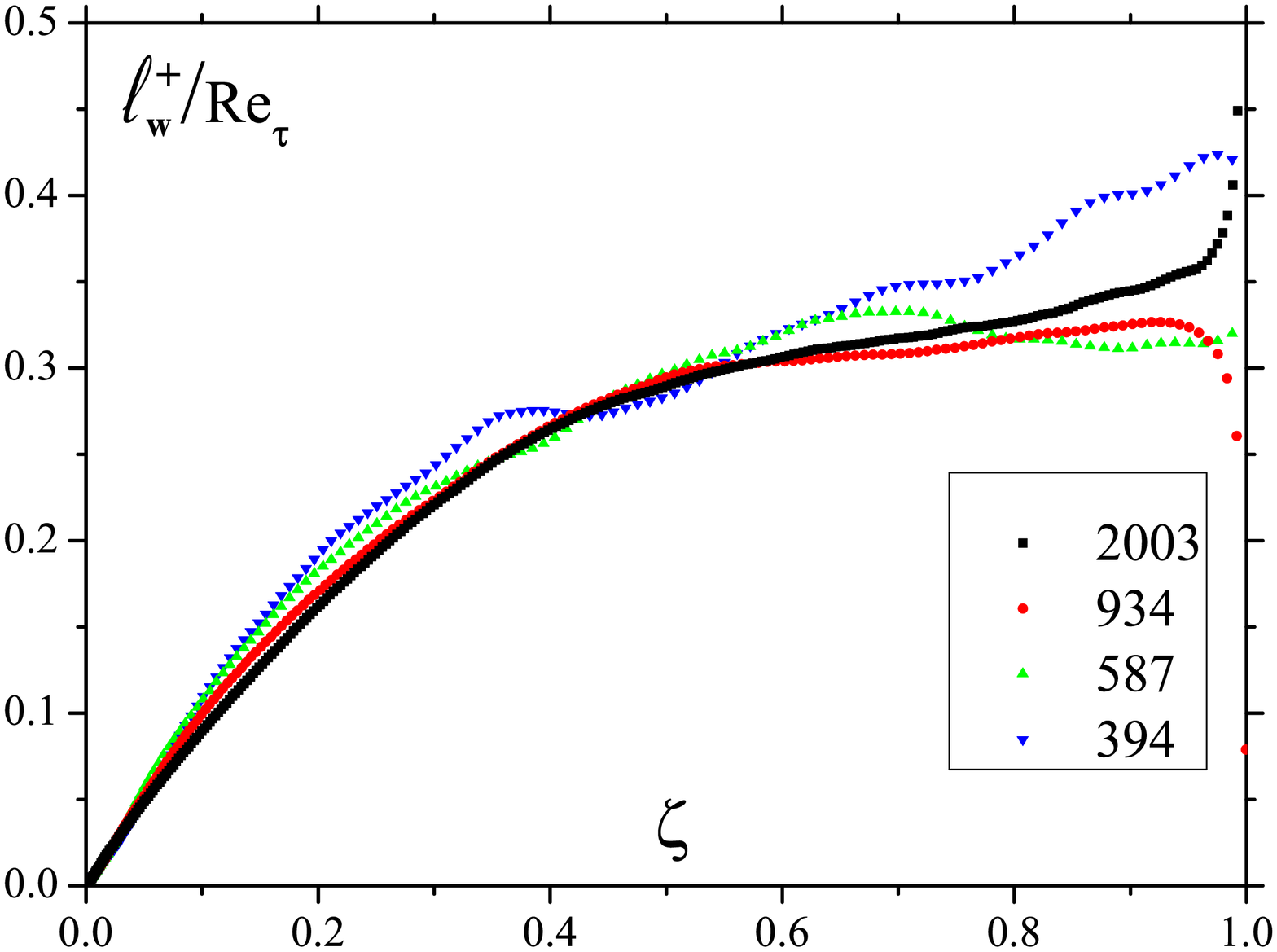}~~
\includegraphics[width=0.485 \textwidth]{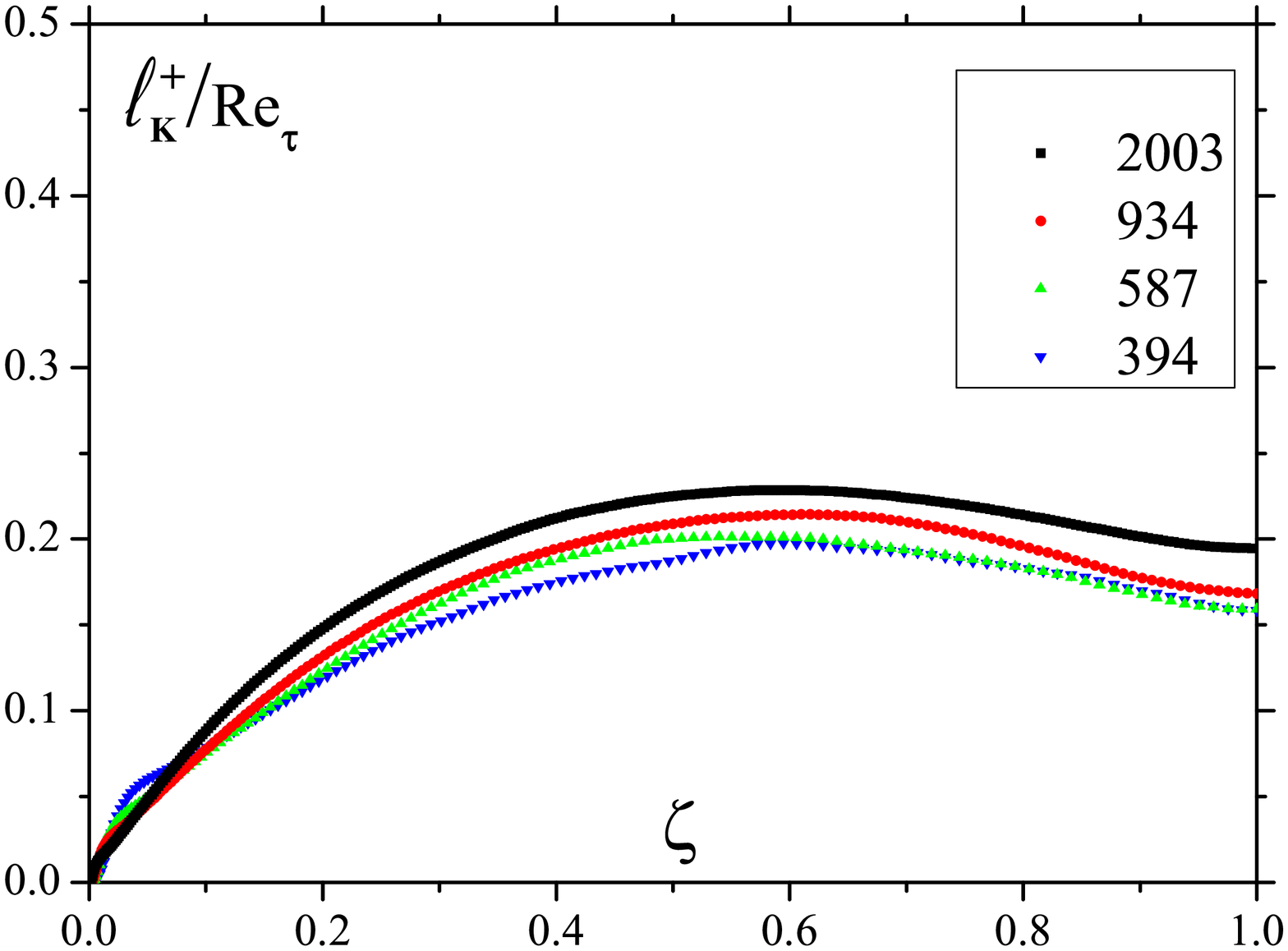}\\ ~\\
\includegraphics[width=0.51 \textwidth]{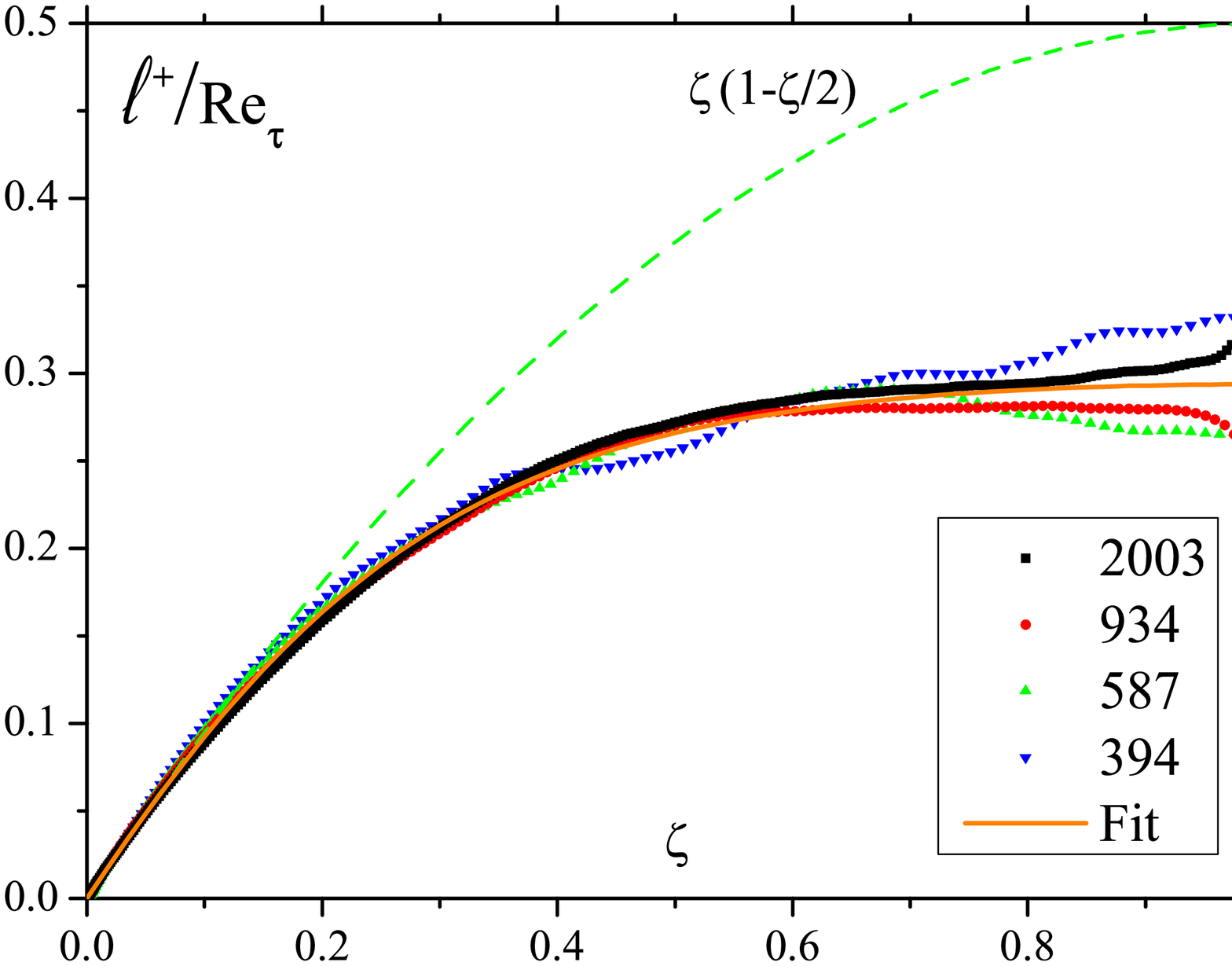}
 \caption{Color online. The scaling function $\ell_{_W}^+(\zeta)/\Ret$ (Left upper panel), $\ell^+_{_K}(\zeta)/\Ret$ (Right upper panel) and the final   scaling function $\ell^+(\zeta)$ (Lower panel), as a function of $\zeta\equiv z/L$, for four different values of $\Ret$, computed from the DNS data~\cite{Moser,DNS}. Note the data collapse everywhere except at $\zeta\to 1$  where $W^+\sim S^+ \ll 1$ and accuracy is lost. The green dash line represents  $\~\zeta=\zeta\,(1-\zeta/2)$ with a saturation level 0.5; in  orange solid line  we show the fitted function
Eq.~(\ref{ell-fit}) with $\ell\sb{sat}=0.311$.}
 \label{ell1}
\end{figure*}
%%%%%%%%%%%%%%%%%%%%%%%%%%%%%%%%%%%%%

A third relation to supplement Eqs.~(\ref{mom}) and (\ref{W}) is obtained by balancing the turbulent energy generated by the mean flow at a rate $SW$, and the dissipation at a rate $\varepsilon_{_K}\equiv \nu \langle |\nabla u|^2\rangle$:
 \begin{equation}
S^+ W^+\approx  \varepsilon^+_{_K}\ ; \quad \varepsilon^+_{_K}={K^+}^{3/2}/[\kappa_{_{ K}}\ell^+_{_{\rm K}}]\ . \label{ene}
 \end{equation}
 Here the dissipation is estimated via the energy cascade over scales involving a characteristic scale of energy containing eddies, $ \ell_{_K} (z)$ determining the energy transfer rate.   The constant $\kappa_{_{ K}}$ will be used to ensure that the slope of this function at $z^+=0$ is unity.

Note that  in \Eqs{W} and~\eq{ene}  we used a local-balance approximation,  neglecting
the spatial energy flux. This approximation is very  good in the log-law region but  it deteriorates near the wall  and near  the center-line. Nevertheless  for our purposes this has no consequences. Near the wall $W^+ \ll S^+$ and the local-balance approximation
plays no role in the exact mechanical balance~\eq{mom} that determines $S$.  For the same reason we also do not need to introduce
a correction  $r_{_ K}(z^+)$ in \Eq{ene}  due to the direct viscous dissipation  (similar to $r_{_ W}(z^+)$  in Eq.~(\ref{W}) since the length
scale replacing $\ell\sb {buf}^+$ here will be the dissipative scale $\ell\sb{diss}\approx 5$ which is entirely buried in the region where  $W$ and $K$ are small. Near the centerline $S^+$ tends to zero and \Eq{mom} determines $W^+\approx 1-\zeta$, which allows an accurate determination of $S^+$, because we know that $\ell_{_W}$ and $\ell_{_K}$ must saturate.

\noindent{\bf Profiles of the characteristic length-scales $\ell_{_K}$, $\ell_{_W}$}: Now  we show that the source of confusion is the assumption that the relevant length scales can be determined {\it a-priori} as   $\ell^+_{_{K,W}}\propto (z^+)^\alpha$ with $\alpha=1$ or $\alpha\ne 1$.   The actual dependence $\ell_{_W}$ and  $\ell_{_K}$ on $z$ and $L$ can  be found from the data
provided by the numerical simulations.  Consider first $\ell_{_W}$, defined by \Eq{W}. We expect that plotting the scaling function
$\ell^+_{_W}/\Ret$ computed for different values of $\Ret$ should collapse the date onto one scaling function. The quality of the data collapse for this scaling function is presented in Fig.\ref{ell1}, demonstrating the expected saturation at the center-line.

The second length-scale, $ \ell^+_{_K}$, is determined by  the second of \Eq{ene}. We again expect that
 $ \ell^+_{_K}/\Ret$ should collapse the data obtained from different value of $\Ret$ onto one scaling function. In Fig.~\ref{ell1}  we demonstrate that this scaling function
leads to acceptable data collapse throughout the channel and for all the four values of $\Ret$ for
which the simulation data are available.
 %%%%%%%%%%%%%%%%%%%%%%%%

 \noindent{\bf Solution, Velocity Profiles and Final Scaling Function}:
Solving Eqs. (\ref{mom}) together with $S^+W^+ = {K^+}^{3/2}/(\kappa_{_{ K}}\ell^+_{_{\rm K}})$ that
follows from Eq. (\ref{ene}),  we find
\begin{equation} W^+ = \big( \kappa \,S^+\ell^+
\big)^2 r_{_W}^{-3/2} \,, \label{solW} \end{equation} where we have
defined the von-K\`arm\`an constant  $ \kappa
\equiv ( \kappa_{_W}^3\kappa_{_K})^{1/4}\approx 0.415$ and the crucial scaling function
$\ell^+(\zeta)$ as follows %%
\begin{equation}\label{defell}
\ell^+ \equiv [{\ell_{_W}^+}^3(\zeta)\, \ell^+_{_K}(\zeta)]^{1/4}  = \sqrt[4]{{W^+}^3r^3_{_W}/{S^+}^3 \varepsilon^+_{_K}} . %%
\end{equation}
Note that if one replaces the energy dissipation rate $\varepsilon^+_{_K}$ by the rate of energy
production $W^+S^+$ and takes $r_{_W}$ as unity this scaling function becomes the Prandtl
mixing length \cite{Pope}. However the latter suffers from a non-physical divergence at the center-line whereas our length saturates to a constant there as it should.

The convincing data collapse for the resulting  function
$\ell^+(\zeta)/\Ret$  is shown in Fig. \ref{ell1}, rightmost panel. Substituting \Eq{solW}
in \Eq{mom} we find
a quadratic equation for $S$ with a solution:
\begin{equation}%%
\label{S+}%%
S^+=\frac{\sqrt{1+{(1-\zeta)[2\kappa
\ell^+(\zeta)}]^2 \big/ r_{_W}(z^+)^{3/2}\,}-1}{2[\kappa\ell^+(\zeta)]^2
\big/ r_{_W}(z^+)^{3/2}} \ .
\end{equation}
To integrate this equation and find the mean velocity profile for
any value of $\Ret$ we need to determine the scaling function
$\ell^+(\zeta)$ from the data.  A careful analysis of the DNS data
allows us to find a good {\it one-parameter} fit for
$\ell^+(\zeta)$
\begin{equation}
\label{ell-fit} \frac{\ell^+(\zeta)}{\Ret} = \ell\sb{s}\Big\{1
-\exp\Big[-\frac{\widetilde{\zeta}}{\ell_s}\,\Big(1
+\frac{\widetilde{\zeta}}{2 \ell_s} \Big)\Big ]\Big\}
\end{equation}
where $ \widetilde{\zeta}  \equiv  \zeta(1-\zeta/2)$ and $
\ell\sb{s} \approx 0.311 $. The quality of the fit is obvious from
the continuous line in the rightmost panel of Fig. \ref{ell1}. Note that the
fit function is exactly constant at mid channel, with zero slope. This is required
by symmetry, and will be the reason for our excellent fit of data in the wake region.
 %%%%%%%%%%%%%%%%

%%%%%%%%%%%%%%%%%%%%

Finally the theory for the mean velocity contains three parameters,
namely $\ell\sb{s}$
together with  $\ell^+\sb{buff}$ (which determines $B$ in Eq. (\ref{loglaw})) and $\kappa$. We demonstrate now that with these three parameters  we can determine the mean velocity
profile for any value $\Ret$, throughout the channel, including the
viscous layer, the buffer sub-layer, the log-law region and the
wake. Examples of the integration of Eq. (\ref{S+}) are shown in
Fig.~\ref{profiles}.  It is worthwhile to re-iterate that the
excellent fits in the viscous and the wake regions (superior to the fits presented in \cite{06LPR,MMI}), which are usually most difficult to achieve,
are obtained here due to the correct asymptotics of $\ell^+(\zeta)$ at $\zeta\to 0$ and $\zeta
\to 1$.  In addition, our theory results also in the kinetic energy, and Reynolds stress profiles which are in a quantitative agreement with the DNS data; for $W$ profiles see Fig.~\ref{profiles}.

%%%%%%%%%%%%%%%%%%%%%
\noindent {\bf Conclusions and application to experiments}: We discussed turbulent channel flow, demonstrating
the existence and usefulness of a scaling function $\ell^+(\zeta)$
which allows us to get the profiles of the mean velocities for all
values of $\Ret$ and throughout the channel, in a good agreement
with DNS.  We argued that the controversy between power-laws and
log-laws is moot, stemming from a rough estimate of the scaling
function $\ell^+(\zeta)$. While
this function begins near the wall as $z^+$, it saturates later, and
its full functional dependence on $\zeta$ is crucial for finding the
correct mean velocity profiles. The approach also allows us to
delineate the accuracy of the log-law presentation, which depends on
$z^+$ and the value of $\Ret$. For asymptotically large $\Ret$ the
region of the log-law can be very large, but nevertheless it breaks
down near the mid channel and near the buffer layer, where
correction to the log-law were presented.

To show that the present approach is quite general, we apply it now
to the experimental data that were at the center of the controversy
\cite{93Bar}, i.e. the Princeton University Superpipe data
\cite{princeton}. In Fig.~\ref{profiles} right panel  we show the mean velocity
profiles as measured in the Superpipe compared with our prediction
using {\it the same scaling function} $\ell^+(\zeta)$. Note that the
data spans values of $\Ret$ from 5050 to 165000, and the fits
with only three $\Ret$-independent constants are very satisfactory.
Note the 2\% difference in the value of $\kappa$ between the DNS and
the experimental data; we do not know at this point whether this
stems from inaccuracies in the DNS or the experimental data, or
whether turbulent flows in different geometries have different
values of $\kappa$. While the latter is theoretically questionable,
we cannot exclude this possibility until a better understanding of
how to compute $\kappa$ from first principles is achieved.

~\\ %%
\textbf{Acknowledgements:}  We thank L. Smits for providing the data of the Princeton Superpipe  and P. Monkewitz and H. Nagib for useful discussion and access to their paper prior to publication.  This
work is supported in part by the US-Israel Binational Science
Foundation.

~\\
\textbf{Appendix: } The exact balance equation for the Reynolds shear stress can be found in \cite{Pope}:
$
   P^+\Sb{W} + \C R^+\Sb{W}= \varepsilon^+\Sb{W} -T^+\Sb{W} \ .
$
Here $P^+\Sb{W} = -\tau^+_{yy} S^+$ is the production of $W^+$,
$R^+\Sb{W}$ is the redistribution of $W^+$ between other Reynolds stress
components, $\varepsilon^+\Sb{W}$ is the viscous dissipation of $W^+$
and $T^+\Sb{W}$ is the turbulent transport of $W^+$. Explicit expressions for
these terms are in \cite{Pope}.  Since $\tau_{yy}$ is $\mathcal{O}(K)$, we approximate
$P^+\Sb{W} \propto - K^+ S^+$. $\C R\Sb{W}^+ = {R\Sb{W}\Sp{\, RI}}^+ +{R\Sb{W}\Sp {\,IP}}^+$
\cite{Pope,06LPR}. The first term describes the return to
isotropy , while the second one is responsible for the
isotropization of production. A slightly modified Rotta's model
\cite{51Rotta} proposes that ${R\Sb{W}\Sp{\, RI}} \propto
\sqrt{K} W/\ell_{_W}$. ${R\Sp {\,IP}}$
is modeled according to \cite{NSW,Pope}, such
that ${R\Sb{W}\Sp {\,IP}} \propto K^+S^+$.

The viscous dissipation
$\varepsilon\Sb{W} \= \nu\<\partial_k u_x \partial_k u_z\>$ is
$\mathcal{O}(-\nu Wz^{-2})$.  As explained in the text, we can
 neglect the non-local term
$T_{_W}$  in the balance for the Reynolds stress with impunity. To compensate for its loss in the viscous range we increase the estimate $(-\nu Wz^{-2})$ by a factor
$\sqrt{K/K_*}$, where $K_*$ is a dimensional constant \cite{06LPR}
Eventually, $\varepsilon^+\Sb{W} \propto -W^+\sqrt{K^+}\big/{z^+}^{2}$.
 Hence, the approximate algebraic balance equation for the Reynolds shear stress reads:
 \begin{equation}\label{W-balance-approx}
   -a K^+S^+ +b \frac{W^+\sqrt{K^+}}{\ell^+_{_W}} +c K^+S^+ \approx -d
\frac{W^+\sqrt{K^+}}{{z^+}^{2}}\,,
 \end{equation}
 where $a, b, c, d$ - are positive constants of $\mathcal{O}(1)$. The last equation may
be rearranged to the form of the fist of Eq. (\ref{W}) but with $
   r_{_W} \=  1 + \ell^+\Sb{buf}\,\ell^+\Sb{W}/{z^+}^2\,, \ \ \
\ \ell^+\Sb{buf} \= d/b$. Since the second term is dominant only near the wall where $\ell^+_{_W}=z^+$, then $ r_{_W} \to  1 + \ell^+\Sb{buf}/z^+$. In \cite{MMI} it was realized that this from, which
is an interpolation between the near wall and the bulk physics, can be modeled
in a way that reflects better the actual width of the buffer layer, using  another interpolation formula that reads
 \begin{equation}\label{r-W-2}
   r_{_W} \=  \Big[1 + \Big(\frac{\ell^+\Sb{buf}}{z^+}\Big)^n\Big]^{1/n}
 \end{equation}
with $n=2$. Best fit to simulational data which is currently available is obtained with $5<n<7$. In this Letter we chose $n=6$ leading to the second of Eqs. (\ref{W}).
This choice simplifies the appearance of the Eqs. (\ref{solW})-(\ref{S+}).

%%  1\\ 2 \\ 3\\ 4 \\ 5 \\ 6\\ 7 \\ 8\\ 9 \\ 10

\end{document}